\documentclass{aims}
\usepackage{amsmath}
  \usepackage{paralist}
  \usepackage{graphics} %% add this and next lines if pictures should be in esp format
  \usepackage{epsfig} %For pictures: screened artwork should be set up with an 85 or 100 line screen
 \usepackage[colorlinks=true]{hyperref}
\hypersetup{urlcolor=blue, citecolor=red}

\def\currentvolume{X}
 \def\currentissue{X}
  \def\currentyear{200X}
   \def\currentmonth{XX}
    \def\ppages{X--XX}

\newtheorem{theorem}{Theorem}[section]
\newtheorem{corollary}{Corollary}

\newtheorem{lemma}[theorem]{Lemma}
\newtheorem{proposition}{Proposition}

\theoremstyle{definition}

\newcommand{\ep}{\varepsilon}
\newcommand{\eps}[1]{{#1}_{\varepsilon}}

\def\eps{\varepsilon}

\def\EE{{\sf E}}
\def\DD{{\sf Var\ }}     %\def\DD{{\sf D}}
\def\NN{{\mathbb N}}
\def\RR{{\mathbb R}}

\title[Nearest-neighbor Entropy Estimators with Weak Metrics]
{Nearest-neighbor Entropy Estimators with Weak Metrics}

\author[Evgeniy Timofeev and Alexei~Kaltchenko]{}


\email{tim@uniyar.ac.ru}
\email{akaltche@wlu.ca}

\thanks{The first author is supported by the Russian government project 11.G34.31.0053.}

\begin{document}
\maketitle
%% Enter the first author's name and address:
\centerline{\scshape Evgeniy Timofeev }
\medskip
{\footnotesize
 %% please put the address of the first author
 \centerline{ Department of Computer Science,  }
   \centerline{Yaroslavl State University, Yaroslavl, 150000, Russia}
} %% Do not forget to end the {\footnotesize by the sign }

\medskip

\centerline{\scshape Alexei~Kaltchenko }
\medskip
{\footnotesize
 %% please put the address of the second author
 \centerline{ Department of Physics and Computing,}
   \centerline{Wilfrid Laurier University, Waterloo, Ontario, N2L3C5, Canada}
} %

\bigskip

% The name of the associate editor will be entered by an editorial staff
% "Communicated by the associate editor name" is not needed for special issue.
 \centerline{(Communicated by the associate editor name)}

%The abstract of your paper
\begin{abstract}
A problem of  improving the accuracy of nonparametric entropy estimation
for a stationary ergodic process is considered. New weak metrics are introduced and  relations between metrics,
measures, and  entropy are discussed.  Based on weak metrics, a new nearest-neighbor entropy  estimator is constructed and has a parameter with which the estimator is optimized to reduce its bias.
It is shown that estimator's variance is upper-bounded by a nearly optimal Cram\'er-Rao lower bound.
\end{abstract}

%The title of your section 1
\section{Introduction}

This study is concerned with improving the accuracy of estimation of the entropy (entropy rate) of
information sources with a finite state space,
whose statistical characterization is unknown.
Sequences of symbols or strings drawn from some finite alphabet appear
in many applications where  objects can be encoded into strings in natural ways.
Such sequences are often  viewed as realizations of stochastic processes also known as  ``information sources''.
An important quantity characterizing an information source is its entropy (entropy rate). For a comprehensive review of previous work on entropy estimation, see, for example, ~\cite{KT08}.
%
%\cite{Sha51,CKV04,Dob58,Gras89,KS94,KASW98,SG96,Ti05mn,WST95} beginning with the Shannon's work~\cite{Sha51}
%on the entropy of English text.
Most widely used are so-called ``nonparametric'' entropy estimators.
However, an analytical evaluation of the accuracy of those estimators is  very difficult and few results are known.
 So, in most published work on nonparametric entropy estimation, only an asymptotical convergence
to entropy is proved and tested  by computer simulation.

For a given  data sample of the size $n$,
the most important characterization  of an estimator~$h_n$  is its efficiency (accuracy), or $L^2$-error
$\EE(h_n - h)^2$,  where~$h$ is the entropy of the source
and~$n$ is the number of observations.
 We recall the relation
$$
\EE(h_n - h)^2 = \DD h_n + (\EE h_n - h)^2,
$$
where the
quantity $\EE(h_n - h)$
 is called \emph{bias}.

Most known nonparametric entropy estimators are based either on Lempel-Ziv compression or the nearest-neighbor method,
 see, for instance, their review in~\cite{KT08}.
 Those estimators are shown to converge almost everywhere(for most general results, see~\cite{KT08}, \cite{KS94}).
 Unfortunately, due to a slow convergence (less then~$O(1/\log n)$), their accuracy is not good  for many practical applications,
 with a relatively  small  sample size of~$\log_2 n$ ($\log_2 n\le 30 - 40$).
 This motivates a search for estimators with a more rapid convergence.

Lempel-Ziv estimators are very hard to analyze and evaluate their accuracy analitically. For example,
the initial motivation for paper~\cite{AS88} was the desire to obtain
asymptotic properties for an entropy estimation algorithm due to
Ziv~\cite{LZ78}, but they show that calculations of the bias and the variance is very
difficult. Up to date, there are no published work on Lempel-Ziv estimator's
bias or variance.

From now on, we will  focus on nearest-neighbor estimators and will now briefly state most important published results. We point out that it is often more convenient to estimate, instead of the entropy, its inverse quantity~$1/h$.

Two modifications of a Grassberger's estimator\cite{Gras89} are proposed in~\cite{Ti05aa}.
In this paper notation, they are written as~$r_n^{(k,m)}(\rho)/\log
n$ (see~\ref{def_r}) and~$\eta_n^{(k,m)}(\rho)$ (see~\ref{etan}),
where~$\rho$ is a metric.

For estimator~$r_n^{(k,m)}(\rho)/\log n$,  $L^1$-convergence and variance bound~$O(n^{-c})$ are shown~\cite{Ti05aa} under certain restriction on source measures. For metric~\ref{rho0} specifically, this measure restriction is relaxed (see \ref{cmu1})   and convergence almost everywhere is established in~\cite{KT08}. It is also shown\cite{KT08} that variance bound~$O(n^{-c})$ holds for any~$c<1$.

For estimator~$\eta_n^{(k,m)}(\rho)$,  $L^1$-convergence is established in~\cite{Ti05aa} under  certain restrictions on metrics and source measures.

For metric~\ref{rho0} computer simulation~\cite{KT08} showed  that the estimator~\ref{etan} with metric~\ref{rho0} is more efficient than the estimator~$r_n^{(k,m)}(\rho)/\log n$. But in a subsequent work~\cite{KT08AEU} for symmetric Bernoulli measures, it was  established  that the estimator's bias  is  a periodic function, with a period proportional to~$\log n$. In a computer simulation, such a bias was difficult to catch because its amplitude was less than~$10^{-6}$ for sources with a small entropy~($h<3$).

In~\cite{Ti11}, the bias was also explicitly calculated for  Markov measures and the metric~\ref{rho0}.
This bias was equal to zero if the logarithms of the transitional probabilities were  rationally incommensurable.
Otherwise,  the bias was a periodic function with a period proportional to~$\log n$. This result demonstrates a
new obstacle  in an estimator's analytical evaluation, namely, an estimator' bias can be a discontinues function of measure parameters.

The objective of this research is to construct a new estimator based on an existing nearest-neighbor estimator
and its modifications  to achieve  efficiency~$O(n^{-c})$ for some measures, where $c>0$ is a constant.
The main idea of this construction is as follows. A nearest-neighbor estimator is based on  some metric.
We introduce a wider class of so-called ``\emph{weak}''\cite{DD09} metrics,
 for which the triangle inequality holds with some constant~$C>1$.
 The new estimator now has a parameter which is a non-decreasing  function.
 We expect that the  function can be selected so that to reduce the bias.
 Specifically, we introduce a class of functions with one parameter which we optimize to reduce the bias.
It is shown that for symmetric Bernoulli measures there exists such a parameter value
for which the bias is asymptotically zero.

Our paper is organized as follows:
\begin{itemize}
  \item
In Section~\ref{Met}, we introduce new weak metrics and discuss a connection between metrics, measures, and  entropy.

  \item
In Section~\ref{NN}, we discuss a nearest-neighbor statistic and show that the statistic's variance is upper-bounded by~$O(n^{-1})$ for a large class of measures and weak metrics,

% In Section~\ref{VAR}
%, which is near the optimal Cram\'er-Rao lower bound.

  \item
In Section~\ref{Est}, we introduce our new nearest-neighbor estimator (based on the statistic of Section~\ref{NN})
and its modifications and prove that this estimator is unbiased for
symmetric Bernoulli measures.

%\item In Section~\ref{Alg}, we provide ideas for estimator's algorithm
%implementation and its computational complexity.
%We also discuss applications of our estimator to dynamical systems.
\end{itemize}

%%%%%%%%%%%%%%%%%%%%%%%%%%%%%%%%%%%%%%%%%%%%%%%%%%%%%%%%%%%%%%%%%%%%%
\section{Notation and Definitions}
\label{Not}
For our purposes, an information source, or stationary process,
is a shift-invariant ergodic measure $\mu$ on
the space $\Omega = A^{\NN}$ of right-sided infinite sequences
drawn from a finite alphabet $A$,
where  $\NN = \{1,2,\dots \} $.
Thus, an infinite random sequence generated by~$\mu$
is viewed as a point in $\Omega$ chosen randomly
with respect to $\mu$ and is denoted by
$\boldsymbol{\xi} = (\xi_1,\xi_2,\ldots)$.

For a cylinder centered at $\boldsymbol{x}\in\Omega$, $s =1,2,\dots$,
we use the following notation
$$C_s(\boldsymbol{x})=\{\boldsymbol{y}\in\Omega : y_1=x_1,\dots, y_s = x_s)\}.$$

Let $\rho$ be a metric on $\Omega$. We denote  an open ball
of radius $r$ centered at $x$ by
$B(\boldsymbol{x},r,\rho)=\{\boldsymbol{y}\in\Omega \ : \
\rho(\boldsymbol{x},\boldsymbol{y})< r\}$. In order to simplify the notation, it is convenient to write
$B(\boldsymbol{x},r)$ for
$B(\boldsymbol{x},r,\rho)$.

Let $\boldsymbol{\xi} = (\xi_1,\xi_2,\ldots)$ be
a point in $\Omega$ chosen randomly with respect to~$\mu$.
Recall
%(see, for instance,~\cite{Bil65})
that the entropy~$h$ (entropy rate)
of a measure~$\mu$ is defined  as follows
\begin{equation}\label{entr}
h = - \lim_{n\to\infty}\dfrac1n\EE\log \mu(C_n(\boldsymbol{\xi})),
\end{equation}
here and throughout the paper, all logarithms are to base $e$, i.e., natural.

\medskip
{\bf Problem  Statement:}\\
Let~$\mu$ be a shift-invariant ergodic probability measure  on~$\Omega = A^{\NN}$.
Let~$\boldsymbol{\xi_0}, \boldsymbol{\xi_1}, \dots, \boldsymbol{\xi_n}$
be independent random variables taking values in $\Omega$ and identically
distributed with a common law $\mu$.  We want to evaluate the entropy of the measure $\mu$.

\section{Metrics on Sequence Spaces}
\label{Met}
Let $\boldsymbol{x} = (x_1,x_2,\dots)$ and
$\boldsymbol{y} = (y_1,y_2,\dots)$ be points in~$\Omega$. We define the following  metric  on~$\Omega$:
\begin{equation}\label{rho}
  \rho(a\boldsymbol{x},b\boldsymbol{y}) =
  \left\{
  \begin{array}{ll}
    e^{-1}\rho(\boldsymbol{x},\boldsymbol{y}), &  a = b;\\
    e^{-\lambda(-\log\rho(\boldsymbol{x},\boldsymbol{y}))}, &  a\ne b;\
  \end{array}
  \right.
  \end{equation}
where  $\lambda(t)$ is a nondecreasing function such that
$\lambda(0)=0$ and $\lambda(t)\le 1$, $0\le t <\infty$.

In particular, for $\lambda(t)=0$, $0\le t <\infty$, we obtain
the following well-known metric:
\begin{equation}\label{rho0}
  \rho_{0}(\boldsymbol{x},\boldsymbol{y}) =
  e^{-\min \left\{k  :   x_{k}\ne y_{k}\right\}},
\end{equation}

We stress that  metric~\ref{rho} is bi-Lipschitz equivalent to
metric~\ref{rho0}, i.e. we have
\begin{equation}\label{bi-L}
\rho_0(\boldsymbol{x},\boldsymbol{y})
\le\rho(\boldsymbol{x},\boldsymbol{y})
\le e\rho_0(\boldsymbol{x},\boldsymbol{y}).
\end{equation}
Therefore, according to~~\cite{DD09}, $\rho$ is a weak metric (or near-metric), i.e. the triangle inequality
holds with some constant $C>1$.

While each point $\boldsymbol{x}$ has infinitely many coordinates,
for any practical estimate calculations,
we need to limit the number of coordinates which are used for calculation.
We make it by introducing a truncation of a metric
that uses only the first~$m$ coordinates of the points.

We define $\rho^{(m)}$, a truncation of a metric~$\rho$, as follows
\begin{equation}\label{cutrho}
  \begin{array}{l}
  \rho^{(0)}(\boldsymbol{x},\boldsymbol{y}) = 1;\\
  \rho^{(m)}(a\boldsymbol{x},b\boldsymbol{y}) =
  \left\{
  \begin{array}{ll}
    e^{-1}\rho^{(m-1)}(\boldsymbol{x},\boldsymbol{y}), &  a = b;\\
    e^{-\lambda(-\log\rho^{(m-1)}(\boldsymbol{x},\boldsymbol{y}))}, &  a\ne b;\
  \end{array}
  \right.
  \end{array}
\end{equation}

To simplify the notation, we define:
\begin{equation}\label{alpha}
\alpha(\boldsymbol{x},\boldsymbol{y})=-\log\rho(\boldsymbol{x},\boldsymbol{y})),\ \
\alpha^{(m)}(\boldsymbol{x},\boldsymbol{y})=-\log\rho^{(m)}(\boldsymbol{x},\boldsymbol{y})).
\end{equation}
Note that $-1/\log\rho_0(\boldsymbol{x},\boldsymbol{y}))$ is also
a  metric on $\Omega$,
but $1/\alpha(\boldsymbol{x},\boldsymbol{y})$ is not a metric.

%!!!!!!!!!!!!!!!!!!!!!!!!!!!!!!!!!!!!!!
\begin{proposition}\label{hdim}
Let~$\mu$ be a shift-invariant ergodic measure on~$\Omega$
and~$\rho$ be metric~\ref{rho};
then, for~$\mu$-almost all points~$\boldsymbol{x}\in \Omega$
$$
\lim_{r\to 0} \dfrac{\log\mu(B(\boldsymbol{x},r,\rho))}{\log r}
= h,
$$
where~$h$ is the entropy of~$\mu$.
\end{proposition}

\begin{proof}
First, we consider a special case~$\rho =\rho_0$.
 Balls in the metric~$\rho_0$ are cylinders; i.e.
$$
  B(\boldsymbol{x},r,\rho_0) = C_n(\boldsymbol{x}), \quad e^{-n-1} < r \le e^{-n}.
$$
Therefore, we have
$$
\lim_{r\to 0} \dfrac{\log\mu(B(\boldsymbol{x},r,\rho_0))}{\log r}
= -\lim_{n\to \infty} \dfrac{\log\mu(C_n(\boldsymbol{x}))}{n}.
$$
Applying
  Shannon--MacMillan--Breiman
theorem~\cite[2.10]{ME84},
we obtain, for~$\mu$-almost all points $\boldsymbol{x}\in \Omega$
$$:
\lim_{r\to 0} \dfrac{\log\mu(B(\boldsymbol{x},r,\rho_0))}{\log r}
= -\lim_{n\to \infty} \dfrac{\log\mu(C_n(\boldsymbol{x}))}{n}
= h.
$$

Now we consider a general metric~\ref{rho}. Since $\rho$ is bi-Lipschitz equivalent(see~\ref{bi-L}) to
a metric~\ref{rho0}, we have
$$
B(\boldsymbol{x},e^{-1}r,\rho_0)  \subset  B(\boldsymbol{x},r,\rho)
\subset B(\boldsymbol{x},er,\rho_0).
$$
Hence, we have
$$
\mu(B(\boldsymbol{x},e^{-1}r,\rho_0))  \le
\mu(B(\boldsymbol{x},r,\rho))
\le \mu(B(\boldsymbol{x},er,\rho_0)).
$$
Therefore, for~$\mu$-almost all points $\boldsymbol{x}\in \Omega$, we obtain:
$$
\lim_{r\to 0} \dfrac{\log\mu(B(\boldsymbol{x},r,\rho))}{\log r}
=h.
$$
\end{proof}

\section{Nearest-neighbor statistics}
\label{NN}

In this section, we consider a nonparametric statistic
$r_n^{(k,m)}(\rho)$.
This statistic is based on a  sample
of~$n+1$ independent points
$\boldsymbol{\xi_{0}}$,  $ \dots$, $\boldsymbol{\xi_{n}}$
in the space~$\Omega$ chosen randomly with respect to~$\mu$
and the metric $\rho$ on~$\Omega$ and
is defined as follows:
\begin{equation} \label{def_r}
 \begin{split}
r_n^{(k,m)}(\rho) =
-\frac{1}{n+1}\sum_{j=0}^{n} \log\left(\min_{i:i \neq j} {^{(k)}}
\rho^{(m)}(\boldsymbol{\xi_{i}}, \boldsymbol{\xi_{j}})\right)
\\
=\frac{1}{n+1}\sum_{j=0}^{n} \max_{i:i \neq j} {^{(k)}}
\alpha^{(m)}(\boldsymbol{\xi_{i}}, \boldsymbol{\xi_{j}}),
 \end{split}
\end{equation}
where $\rho$ is a metric~\ref{rho}
and $\min ^{(k)}\{X_1,\dots,X_N\}$ is defined by  $\min ^{(k)}\{X_1,\dots,X_N\} = X_k$ if
$X_1\le X_2\le \dots\le X_N$.

We stress that this statistic uses only first $m$ coordinates
of  points $\boldsymbol{\xi_{0}}$,  $ \dots$, $\boldsymbol{\xi_{n}}$.

%%%%%%%%%%%%%%%%%%%%%%%%%%%%%%%%%%%%%%%%%%%%%%%%%%%%%%%%%%%%%%%%%%%%%%%

Theorem 1 and Proposition 8 of~\cite{Ti05aa} imply the following
statement:
\begin{proposition}\label{EDim}
Let
$\boldsymbol{\xi_{0}}$,  $ \dots$, $\boldsymbol{\xi_{n}}$
be~$n+1$ independent points in the space~$\Omega$
chosen randomly with respect to~$\mu$ and $k = O(\log n)$,
then the following limit holds:
$$
\lim_{n\to\infty}\frac{\EE r_n^{(k,\infty)}(\rho)}{\log n}
= \frac1h .
$$
\end{proposition}

%%%%%%%%%%%%%%%%%%%%%%%%%%%%%%%%%%%%%%

\begin{lemma}\label{mEr}
Let a measure~$\mu$ satisfy
the following condition
\begin{equation}\label{cmu1}
\exists a, b >0 \ : \  \mu(C_n(\boldsymbol{x},r)) \le be^{-an},\
\ \forall n>0, a.e. \boldsymbol{x}\in\Omega,
\end{equation}
then,
there exist  constants $c_1, c_2$ such that
the following inequality holds:
\begin{equation}\label{dErm}
 \EE r^{(k,\infty)}_{n}(\rho) - \EE r^{(k,m)}_{n}(\rho) \le c_1n^{-1},\ for \  m \ge c_2\log n.
\end{equation}
\end{lemma}
\begin{proof}
Arguing as in proof
of Theorem 1~\cite{Ti05aa}, we obtain
\begin{equation}\label{Er_n}
 \begin{split}
\EE r_{n}^{(k,m)}(\rho)
 =
-n\binom{n-1}{k-1}
\int_{\Omega}\int_{0}^{1} \log r\,\mu(B(\boldsymbol{x},r,\rho^{(m)}))^{k-1}
\\
\left(1-\mu(B(\boldsymbol{x},r,\rho^{(m)}))\right)^{n-k}
d_r\mu(B(\boldsymbol{x},r,\rho^{(m)}))d\mu(\boldsymbol{x}).
 \end{split}
\end{equation}
%\EE r_{n}^{(k,m)} =n\binom{n-1}{k-1}\int_{0}^{1} \int_{\Omega}\log\nu^{(m)}(t,\boldsymbol{x},\rho)t^{k-1}(1-t)^{n-k}d\mu(\boldsymbol{x})dt.
From an identity $$
B(\boldsymbol{x},r,\rho) = B(\boldsymbol{x},r,\rho^{(m)}), \ r >
e^{-m},
$$
we get
\begin{equation*}
 \begin{split}
\EE r_{n}^{(k,\infty)}(\rho)
-\EE r_{n}^{(k,m)}(\rho)
 =
-n\binom{n-1}{k-1}
\int_{\Omega}\int_{0}^{e^{-m}} \log r\,\mu(B(\boldsymbol{x},r,\rho))^{k-1}
\\
\left(1-\mu(B(\boldsymbol{x},r,\rho))\right)^{n-k}
d_r\mu(B(\boldsymbol{x},r,\rho))d\mu(\boldsymbol{x}).
 \end{split}
\end{equation*}
Using Condition~\ref{cmu1} and~\ref{bi-L}, we obtain
$$
\mu(B(\boldsymbol{x},r,\rho)) \le c_3r^a.
$$
Therefore, we have
$$
\EE r_{n}^{(k,\infty)}(\rho)
-\EE r_{n}^{(k,m)}(\rho)
 \le
-n^{k}
\int_{0}^{e^{-m}} \log r\,(c_3r^a)^{k-1}\,dr.
$$
Calculating the above integral, we obtain
$$
\EE r_{n}^{(k,\infty)}(\rho)
-\EE r_{n}^{(k,m)}(\rho)
= O\left(n^{k}m e^{-m(a(k-1)+1)}\right).
$$
If we set~$c_2 > 1/a$, then the inequality~\ref{dErm} follows from the above equality.
\end{proof}
Applying Proposition~\ref{EDim}, we get the following statement:
\begin{corollary}\label{lERnm}
For some constant~$c > 1/a$, let the following relations hold:
$$c\log n \le m , \ \ k = O(\log n),$$
then the following limit holds:
$$
\lim_{n\to\infty}\frac{\EE r_n^{(k,m)}}{\log n}
= \frac{1}{h}.
$$
\end{corollary}

\begin{theorem}\label{Dr}
Let $r^{(k,m)}_{n}$ be a statistic defined in~(\ref{def_r}),
then the following inequality holds
\begin{equation}\label{Dr_n}
\DD r_n^{(k,m)}
 \le \frac{m^2 (km+1)^2 }{4(n+1)}.
 \end{equation}
\end{theorem}
\begin{proof}
In this proof, we use McDiarmid's method \cite{McD89}.

We introduce a function $$f \ : \ \Omega^{n+1} \to \RR$$ defined as
$$
f(\boldsymbol{x_0},\dots,\boldsymbol{x_n})
=
%\triangle
\frac{1}{n+1}\sum_{j=0}^{n} \max_{i : i\ne j} {^{(k)}}
\alpha^{(m)}(\boldsymbol{x_i}, \boldsymbol{x_j}).
$$

In order to apply McDiarmid's method, we need to show that $f$
satisfies the inequality
\begin{equation} \label{bdMcD}
\sup_{\boldsymbol{x_0},\dots,\boldsymbol{x_n},\boldsymbol{y} \in\Omega}
 |f(\boldsymbol{x_0},\dots,\boldsymbol{x_n})-
f(\boldsymbol{x_0}\dots,\boldsymbol{x_{i-1}},\boldsymbol{y},
\boldsymbol{x_{i+1}},\dots,\boldsymbol{x_n})| \le c_i,
\end{equation}
for all $0 \le i \le n$.

We prove this inequality for
\begin{equation}\label{ciMcD}
  c_i = \frac{m(km+1)}{n+1} , \ \ 0\le i \le n.
\end{equation}
For brevity, we introduce the following notation
\begin{align*}
 \boldsymbol{X}= (\boldsymbol{x_0},\dots,\boldsymbol{x_n}),\\
\boldsymbol{\tilde X} = (\boldsymbol{x_0},\dots,\boldsymbol{x_{i-1}},y,
\boldsymbol{x_{i+1}},\dots,\boldsymbol{x_n}),\\
g_j(\boldsymbol{X}) =
\max_{i :i\ne j} {^{(k)}} \alpha^{(m)}(\boldsymbol{x_i}, \boldsymbol{x_j}).
\end{align*}

Let
$
J =\left\{ j\ne i\ : \ g_j(\boldsymbol{X}) \ne g_j(\boldsymbol{\tilde X})
\right\}.
$

Since
\begin{align*}
g_j(\boldsymbol{X}) = g_j(\boldsymbol{\tilde X}), j\notin J, \ j \ne i;\\
|g_j(\boldsymbol{X}) - g_j(\boldsymbol{\tilde X})| \le m, \forall j;
\end{align*}
we have
$$
|f(\boldsymbol{X})- f(\boldsymbol{\tilde X})| \le \frac{m(|J|+1)}{n+1}.
$$

Let us prove that $|J| \le km$.

If  $j\in J$ then $g_j(\boldsymbol{X}) =
\alpha^{(m)}(\boldsymbol{x_i}, \boldsymbol{x_j})$.

Suppose that $l = 1, 2,\dots, m$ is such that
$$
\boldsymbol{x_j}\notin C_l(\boldsymbol{x_i}), \ \
\boldsymbol{x_j}\in C_{l-1}(\boldsymbol{x_i}),
$$
then $\forall \boldsymbol{y}\in C_l(\boldsymbol{x_i})$
$$
\alpha^{(m)}(\boldsymbol{y}, \boldsymbol{x_j}) > \alpha^{(m)}(\boldsymbol{y},
\boldsymbol{x_i}).
$$
Hence,
$$
\left|C_l(\boldsymbol{x_i})\bigcap \{\boldsymbol{x_{0}},  \dots,
\boldsymbol{x_{n}}\}\right|\le k.
$$

This proves~(\ref{ciMcD}).

As shown in~\cite{Dev91}, McDiarmid's martingale method
provides the following bound on the variance of $f(\boldsymbol{\xi_0},
\dots,\boldsymbol{\xi_n})$
\begin{equation}\label{L2zeta}
\DD\left[f(\boldsymbol{\xi_0},\dots,\boldsymbol{\xi_n})\right]
\le \frac14\sum_{a=0}^n c_a^2.
\end{equation}
Substituting~(\ref{ciMcD}) into~(\ref{L2zeta}), we obtain~(\ref{Dr_n}).

\end{proof}

\begin{corollary}\label{con}
\begin{equation}\label{Mcbias}
P\{| r_n^{(k,m)}(\rho) - \EE r_n^{(k,m)}(\rho)|>\delta\}
\le 2e^{-2(n+1)\delta^2/m^2(km+1)^2}.
\end{equation}%
\end{corollary}

\begin{proof}
This inequality is obtained by applying the McDiarmid's inequality~\cite{McD89}
$$
%\begin{multline*}
P\{| f(\boldsymbol{\xi_0},\dots,\boldsymbol{\xi_n}) -
\EE f(\boldsymbol{\xi_0},\dots,\boldsymbol{\xi_n})|>\delta\}
\le 2e^{-2\delta^2/\sum_{i=0}^n c_i^2},
$$
%\end{multline*}
where $c_i$ is defined in~(\ref{bdMcD}).

Substituting~(\ref{ciMcD}) into the above inequality,  we get
$$
P\{| f(\boldsymbol{\xi_0},\dots,\boldsymbol{\xi_n})
 - \EE f(\boldsymbol{\xi_0},\dots,\boldsymbol{\xi_n})|>\delta\}
\le 2e^{-2(n+1)\delta^2/m^2(km+1)^2}.
$$
Substituting $r_n^{(k,m)}(\rho) =f(\boldsymbol{\xi_0},\dots,\boldsymbol{\xi_n})$, we get~\ref{Mcbias}.
\end{proof}
%

%%%%%%%%%%%%%%%%%%%%%%%%%%%%%%%%%%%%%%%%%%%%%%%%%%%%%%%%%%%%%%%%%%%%%%%%%%%%%%%%%%%%%%%%%%%%%%%%%%%%%%%%%%%%%%%%
\begin{corollary}\label{C_rho01}
Let a measure~$\mu$ satisfy~\ref{cmu1} and an inequality~ $c\log n \le m $ hold for some constant $c > 1/a$, $m = O(\log n)$, and $k = O(\log n)$,
then a sequence
$\dfrac{\EE r_n^{(k,m)}(\rho)}{\log n}$ converges  to~$1/h$ a.e.
\end{corollary}
%%%%%%%%%%%%%%%%%%%%%%%%%%%%%%%%%%%%%%%%%%%%%%%%%%%%%%%%%%%%%%%%%%%%%%%%%%%%%%%%%%%%%%%%%%%%%%%%%%%%%%%%%%%%%%%%

\section{Entropy Estimator}\label{Est}
In this section, we consider a nonparametric estimator~$\eta_n^{(k,m)}(\rho)$ for
the inverse entropy  $1/h$,
where metric~$\rho$ is defined in~\ref{rho}.

 This estimator is based on a  sample
of~$n+1$ independent points
$\boldsymbol{\xi_{0}}$,  $ \dots$, $\boldsymbol{\xi_{n}}$
in the space~$\Omega$ chosen randomly with respect to~$\mu$
and the metric~\ref{rho} on~$\Omega$
and is defined as follows:
\begin{equation} \label{etan}
\eta_n^{(k,m)}(\rho) = k \left(r_{n}^{(k,m)}(\rho) - r_{n}^{(k+1,m)(\rho)}\right),
\end{equation}
where
$r_n^{(k,m)}(\rho)$ is defined in~\ref{def_r}.
%%%%%%%%%%%%%%%%%%%%%%%%%%%%%%%%%%%%%%%%%%%%%%%%%%%%%%%%%%%%%%%%%%

Applying Theorem~\ref{Dr} and inequality
$\DD(X+Y) \le \left(\sqrt{\DD X} + \sqrt{\DD Y}\right)^2$, we obtain the following statement:

\begin{proposition}\label{Deta}
Let
$m = O(\log n)$ and $k = O(\log n)$,
then
$$
\DD \eta_n^{(k,m)}(\rho) = O(n^{-1} \log^8 n).
$$
\end{proposition}
\noindent Thus, we have just calculated the estimator's variance. We note that a calculation of the estimator's bias is much more complicated and we will do it for a asymmetric Bernoulli measure.
\begin{proposition}\label{symBer}
Let~$\mu$ be a asymmetric Bernoulli measure and the function~$\lambda(t)$ of~\ref{rho} be such that the following identity holds
\begin{equation} \label{lambda1}
  \lambda(t) = \log_{\beta}(\beta + (1-\beta)\beta^t)
  \end{equation} for~$0<\beta < 1$.

Then, for $\beta = 1/|A|$, we have
$$
\EE\eta_n^{(k,\infty)}(\beta) = 1/h = 1/\log |A|.
$$
\end{proposition}

\begin{proof}
We introduce notation 
$$
F(t) = \mu(B(\boldsymbol{x},e^{-t})).
$$
Clearly, for a symmetric Bernoulli measure (with equiprobable
symbols), $F$ does not depend on $\boldsymbol{x}$ and satisfies an
equation
$$
F(t) = \beta F(t-1) +(1-\beta)F(\lambda^{-1}(t)),
$$
where $\beta = 1/|A|$.
It can easily be checked that $F(t) =\beta^t$ is a solution of this
equation.

Substituting $F(t) =\beta^t$ for $m=\infty$ in~\ref{Er_n}, we get
$$
\EE r_{n}^{(k,\infty)}(\rho)
 =
-n\binom{n-1}{k-1}
\int_{0}^{1} \log r\,\beta^{-(k-1)\log r}
\left(1-\beta^{-\log r}\right)^{n-k}
d\beta^{-\log r}.
$$
If we replace $ \beta^{-\log r}$ by $x$, we obtain
$$
\EE r_{n}^{(k,\infty)}(\rho)
 =
\frac{n}{\log\beta}\binom{n-1}{k-1}
\int_{0}^{1} \log x\,x^{(k-1)}
(1-x)^{n-k}
dx.
$$
Calculating the integral (see~\cite[4.253.1]{GR71}), we get
$$
\EE r_{n}^{(k,\infty)}(\rho)
 =
-\frac{H_n-H_{k-1}}{\log\beta},
$$
where $H_n$ are harmonic numbers
$$
H_n = \sum_{s=1}^n\frac1s.
$$
Using~\ref{etan}, we obtain
$$
\EE \eta_{n}^{(k,\infty)}(\rho)
 =
-\frac{1}{\log\beta}.
$$
\end{proof}

\section{Conclusion}

In this work, we have introduced a new nearest-neighbor entropy  estimator which is based on a new large family of weak metrics. The estimator has a parameter with which it is optimized to reduce its bias. We have calculated estimator's variance and shown that it is upper-bounded by a nearly optimal Cram\'er-Rao lower bound.

We have explicitly calculated the estimator's bias for  a special case - symmetrical Bernoulli measures. In a subsequent work, we expect to calculate the bias for a general case as well as to develop an efficient estimator's algorithm implementation.

\end{document}